\documentclass[12pt]{article}
\newlength{\dinwidth}
\newlength{\dinmargin}
\begin{document}
\font\blackboard=msbm10 at 12pt
 \font\blackboards=msbm7
 \font\blackboardss=msbm5
 \newfam\black
 \textfont\black=\blackboard
 \scriptfont\black=\blackboards
 \scriptscriptfont\black=\blackboardss
 \def\bb#1{{\fam\black\relax#1}}
\newcommand{\z}{{\bb Z}}
\newcommand{\real}{{\bb R}}
\newcommand{\bc}{{\bb C}}
\newcommand{\be}{\begin{equation}}
\newcommand{\ee}{\end{equation}}
\newcommand{\ber}{\begin{eqnarray}}
\newcommand{\eer}{\end{eqnarray}}
\newcommand{\lp}{\left(}
\newcommand{\rp}{\right)}
\newcommand{\lk}{\left\{}
\newcommand{\rk}{\right\}}
\newcommand{\lc}{\left[}
\newcommand{\rc}{\right]}
\newcommand{\sT}{{\scriptscriptstyle T}}
\newcommand{\2}{\,\,2}
\newcommand{\se}{\section}
\newcommand{\tra}{\vec{p}_{\sT}}
\newcommand{\Z}{Z\left(\beta\right)}
\newcommand{\half}{\frac{1}{2}}
\renewcommand\theequation{\thesection.\arabic{equation}}
\thispagestyle{empty}

\begin{flushright}
\begin{tabular}{l}
{\tt hep-th/0011178}\\
\end{tabular}
\end{flushright}

\vspace*{2cm}

{\vbox{\centerline{{\Large{\bf On the effective character of a non abelian
DBI action
}}}}}

\vskip30pt

\centerline{M. A. R. Osorio and Mar\'{\i}a Su\'arez\footnote{E-mail
addresses: osorio, maria@string1.ciencias.uniovi.es}}

\vskip18pt

\begin{center}

{\it Dpto. de F\'{\i}sica \\ Universidad de Oviedo \\
Avda. Calvo Sotelo 18\\ E-33007 Oviedo, Asturias, Spain}

\end{center}

\vskip .5in

\begin{center}
{\bf Abstract} 
\end{center}
We study the way Lorentz covariance can be reconstructed from Matrix Theory
as a IMF description of M-theory.  The problem is actually related to the
interplay between a non abelian Dirac-Born-Infeld action and Super-Yang-Mills
as its generalized non-relativistic approximation. All this physics shows up
by means of an analysis of the asymptotic expansion of the Bessel functions
$K_\nu$ that profusely appear in the computations of amplitudes at finite
temperature and solitonic calculations. We hope this might help to better
understand the issue of getting a Lorentz covariant formulation in relation
with the $N\rightarrow +\infty$ limit. There are also some computations that
could be of some interest in Relativistic Statistical Mechanics.

\newpage

\section{Introduction}

It is widely admitted that  Matrix Theory is a non perturbative formulation
of eleven dimensional M-theory either as its discrete light cone
quantization (DLCQ) or in the infinite momentum frame (IMF) in the limit in
which one connects with the perturbative superstring and its related objects
as Dp-branes (see for example \cite{matrix} and references therein).

What we call Matrix Theory is the Quantum Mechanics which results from the
dimensional reduction of ${\cal N} = 1$ SYM theory in ten dimensions to
$0+1$ dimensions. As a system in one dimension it has eight target degrees
of freedom of bosonic and fermionic kind. With them one can finally describe
the 256 (=128+128) degrees of freedom ${\cal N}= 1$ supergravity has in
eleven dimensions. A light-cone and an IMF description of a theory in $D$
dimensions is actually a theory in $D-2$ dimensions. Eight are the number of
degrees of freedom that result from the gauge symmetries the SYM has. In the
Hamiltonian formalism of this quantum mechanics, it is easy to see that the
extra Gauss constraints that reduce the number of SYM gauge fields to eight
become trivial when the gauge group becomes abelian as actually happens in
the limit in which the $N$ branes get apart. This is the free limit which
corresponds to the low energy limit of M-theory in which massless
${\cal N}=1$ SUGRA in eleven dimensions is recovered. And so one finally
finds oneself with nine degrees of freedom which are the number of
transverse dimensions in a eleven dimensional space-time.

If from the very beginning one puts the system at finite temperature, the
DLCQ of M-theory that Matrix Theory provides easily produces a canonical
free energy that, in the classical limit when the radius $R_+$ of $x^-$ goes
to infinity, gives the Helmholtz free energy of the corresponding massless
SUGRA \cite{ours}. This is an expected result because a generic observer is
linked to the light-cone frame by a (limiting) Lorentz transformation
\cite{Seiberg-sen}.

What we will do is to study the relationship between the IMF calculation of
the free energy and the same magnitude gotten for massless
${\cal N}=1$ SUGRA in a generic Lorentz frame (what loosely could be called
the 'Lorentz covariant' version of that amplitude). This will reveal us the
way IMF computations can be connected with generic frame descriptions. In
particular, we will learn, in connection with the Matrix Theory conjecture,
on the physical meaning of the Bessel functions $K_{\nu}$ which profusely
appear in finite temperature and solitonic calculations (see \cite{Green}
for the special case of D-instanton calculations) and, in particular, their
asymptotic expansions on the variable $\beta m$. We will actually see that
such expansion is one coming from the Galilean average of the relativistic
corrections of the energy of the KK modes in the uncompactified ten
dimensional space-time. This way we will show that the expansion directly
results from one for the exponential of the relativistic energy of a
Kaluza-Klein-particle of mass $m_k$ in ten dimensions as a power series in
$\lp v_T/c\rp^2$ where $c$ is the speed of light and $v_T$ the modulus of
the nine dimensional transverse velocity. The asymptotic character will be a
result of Watson's Lemma because taking the trace over the nine open
dimensions (which amounts in fact to a Galilean average) will be equivalent
to making a Laplace transform of the power series in $\lp v_T/c\rp^2$. The
IMF limit keeps, from this series, the rest energy plus the Galilean
contribution. This tells us that there must be a difference in the way we
catch the eleven dimensional degrees of freedom if we compare the IMF
approach and the DLCQ picture. In particular, in the IMF picture it will be
relevant whether those degrees of freedom can be captured without an
infinite number of anti-D0-branes. The anti-D0-branes are explicit in the
generic description in which negative values of $p_{11}$ appear. We will
actually show how all the degrees of freedom of the uncompactified SUGRA can
be gotten by using only positively RR-charged states. After all, DLCQ and
the IMF description have to coincide with any generic description when we
take the decompactification limit
\cite{9704080}.

From here, it will be easy to construct the effective Lagrangian of the
(free) single-object as seen by a generic observer. That is to say, the
D-particle description of the classical limit of M-theory. It will be none
other than the Dirac-Born-Infeld (DBI for short) Lagrangian for U(1)$^k$. 
It would be a limit of the one for $U(k)$ defined as some still unknown
generalization of the abelian DBI Lagrangian. It is worth remarking that the
trace would always be performed through an expansion in powers of matrices
taken as a definition for the square root of a non diagonal matrix. That
would finally produce an asymptotic series for the partition function after
making an expansion for the exponential.  In other words, the effective
character of any non abelian generalization of the DBI action would show in
which, through Watson's Lemma, one would get an asymptotic expansion for the
canonical free energy. We will see that to actually get the complete number
of degrees of freedom after decompactifying to eleven dimensions, a more
subtle view on the extension through analytic continuation of the effective
character of this expansion is needed.

First of all, let us recall in section two the finite temperature classical
limit of M-theory as seen in a generic Lorentz frame. After that, we will
present the same problem from the IMF view. Section four will explain how to
relate the asymptotic expansion with any eventual generalization of the DBI
action for the non abelian $U(k)$ case and will establish the conclusions too.

\section{The free energy as seen by a generic observer}

The single-object  contribution to the canonical partition function for
${\cal N}=1$, $d=11$ SUGRA on $S^1\times {\bb R}^{10}$ as calculated in a
generic frame admits a proper time representation which reads

\be
Z_1[\beta] =-\beta F_{MB}[\beta] = 256 V_{10} 
\beta\lp 2\pi\rp^{-5} \int_{0}^{+\infty}\, ds \,
s^{-13/2}\,\theta_3\lc 0,\frac{\mbox{i}\left(2 \pi R_{11}\right)^2}{2\pi s}\rc
\,e^{-\beta^2/2s}
\label{proper}
\ee
where $F_{MB}$ stands for the Maxwell-Boltzmann contribution to the Helmholtz
free energy.

As the classical limit of M-theory on ${\bb R}^{10}\times S^1$, this is the
result of taking the limit $\sqrt{\alpha'} \ll (\sqrt{\alpha'} g_s)$ with
$\alpha'$ going to zero and $R_{11}= g_s \sqrt{\alpha'}$, cf.
\cite{chicos}. This becomes more explicit after performing a Poisson
re-summation over the Jacobi $\theta_3$ function to see physics as
pictured in ten dimensions, namely

\be
Z_1[\beta] = 256 V_9 \beta \lp 2\pi\rp^{-5} \int_{0}^{+\infty}\, ds \,
s^{-6}\,\lp 2\,\sum_{k=1}^{+\infty} e^{-m_k^2 s/2} \,e^{-\beta^2/2s}\,
+ \,e^{-\beta^2/2s}\rp
\label{properF}
\ee 
where the last term in the integrand gives the $\alpha'\rightarrow 0$ limit
of the contribution of the free energy of the SST IIA which corresponds to
the zero mode along the compact eleventh dimension. The sum over KK modes
has been trailed to a sum on
$m_k=k/R_{11}$ from one to infinity picking up a factor of two. This
means that D0-branes (positive momentum) and anti-D0-branes (negative
momentum) contribute just the same.

By performing the proper time integral one obtains

\be
Z_{1}[\beta] = (k=0\,\mbox{contribution}) + 
2\beta^{-4}V_9\sum_{k=1}^{+\infty}\,2 
\lp 2\pi\rp^{-5}m_k^5\,K_5\lc \beta m_k\rc
\label{Besselg}
\ee
Here, $K_5$ is a modified Bessel function. It admits an asymptotic expansion
which is

\be
K_5\lc \beta m_k\rc = \sqrt{\frac{\pi}{2\beta m_k}}\,
e^{-\beta m_k}\sum_{n=0}^{+\infty}\frac{\Gamma\lc n + 11/2\rc}{\Gamma\lc
n+1\rc\Gamma\lc -n+11/2\rc}\lp 2\beta m_k\rp^{-n}
\label{asymg}
\ee
Let us now see  the IMF description we get for this limit of M-theory. In
there, it is assumed that Super-Yang-Mills is the Galilean approximation of
a Lagrangian that covariantly would read as a non abelian generalization of
the DBI action.

\section{Finite temperature (low energy) M-theory as seen in the IMF}

An infinite momentum frame description of M-theory is one in which the total
momentum in the longitudinal direction is much larger than all transverse
momenta. The connection between the IMF frame and the traditional light-cone
frame is provided by the fact that the momentum in the longitudinal
direction on which the observer is boosted satisfies $p_L\gg p_{-}$. The
M(atrix) theory proposal in its IMF version assumes that a covariant version
would be achieved after adding up the relativistic corrections to the
Galilean part.

The connection between the IMF description and that in a generic Lorentz
frame is based on the fact that in the IMF limit where $p_{11}\gg p_{T}$ we
have
\be
p^0 = \sqrt{p_{11}^2 + \vec{p}_{\sT}^{\,\,2}} = |p_{11}| \left\{1+
\frac{{p_{\sT}}^2}{2|p_{11}|^2} + {\cal O}
\left[\left({p_{\sT}^2\over p_{11}^2}\right)^{2}
\right]\right\}
\label{desarrollo}
\ee
This is the expression of  the energy of a free massless particle in eleven
dimensions. This will be the Hamiltonian of the relativistic
Supersymmetric Quantum Mechanics of the KK-modes with mass $|p_{11}|$ in
ten dimensions, i.e., the Hamiltonian for a free bound state of D0-branes. 
Now, since $x^{11}$ is compact, $p_{11}$ will be quantized in units of
${1/R_{11}}$. 

Nonetheless, what we want to do is to compute $Z_1(\beta)$ from the IMF by
introducing the relativistic corrections. Instead of using the expression of
the energy as the square root, we write
$p^0$ as the IMF approximation plus the corrections. Restoring the speed of
light $c$ for the moment, and writing $m_k$ for the mass of the
KK mode, $m_k = {k\over cR_{11}}$, we find
\be
{p^0\over m_{k}c}= 1 + {p_{\sT}^2\over 2 m_k^2 c^2}+ 
{\cal C}\left({p_{T}^2\over m_k^2 c^2}\right)
\ee
where ${\cal C}[x]$ can be represented  by  the series
\be
{\cal C}[x]={\lp 1/2\rp !}
\sum_{n=1}^{\infty}{x^{n+1}\over (-1/2 - n)!(n+1)!}
\label{seriee}
\ee
that, to directly compute several orders,  can be used as a fast way of
reading the $n$-th derivative of ${\cal C}[x]$ at $x=0$.

All together will provide us with an expansion in powers of
$p_{\sT}^{2}/(m^2 c^2)$ for $\mbox{e}^{-\beta c p^0}$ that 
is\footnote{We would like to emphasize the fact that it is an expansion for
the exponential of the energy times the inverse temperature and not
necessarily for the square root},
\be
e^{-\beta\,c\,p^0} = 
e^{-\beta m_k c^2}e^{-\beta  p_{\sT}^2/2m_k}\,\sum_{n=0}^{+\infty}\,
a_{n}\lc\beta m_{k}c^2\rc\left({p_{\sT}^2\over m_k^2 c^2}\right)^{n}
\label{t}
\ee
where 
\be
a_{n}[y]=\frac{1}{n!}
\frac{d^n}{dx^{n}}\left. e^{-y\,{\cal C}[x]}\right|_{x=0}
\label{coefa}
\ee
will only depend on dimensionful quantities through the 
dimensionless combination $y\equiv\beta m_{k}c^2$ and are, in fact, finite
degree polynomials on this variable. It is remarkable that we
are keeping apart the Boltzmann weight corresponding to the rest  plus
the Galilean kinetic energy. This factorization will actually be taken as
the IMF way of computing the partition function. For instance, 
neglecting terms of ${\cal O}(v_T^{10}/c^{10})$ we have
\begin{eqnarray}
e^{-\beta\,c\, p^0} &=&
 e^{-\beta m_k c^2}{e}^{-\beta p_\sT^2/2m_k}\,\left\{ 1 +{1\over 8} (\beta
m_kc^2)\left({p_{T}\over m_k c}\right)^4- {1\over 16}(\beta
m_kc^2)\left({p_{T}\over m_k c}\right)^6
\right. \nonumber \\  &+& \left. 
{1\over 128}(\beta m_kc^2)(\beta m_kc^2+5)\left({p_{T}\over m_k c}\right)^8
+{\cal O}\left[\left({p_{T}\over m_k c}\right)^{10}\right]\right\}
\end{eqnarray}

We can now compute the partition function by taking the trace over
transverse  momenta an summing over $k$ as well. For the transverse
part we find from (\ref{t}) 
\be Z_{T}[\beta]={\rm Tr}_{\vec{p}_{\sT}}e^{-\beta
cp^0}= e^{-\beta m_k c^2}\sum_{n=0}^{+\infty}\, {a}_{n}\lc\beta
m_{k}c^2\rc\left\langle\left({p_{T}\over m_k c}\right)^{2n}
\right\rangle_{\beta} 
\ee 
where $\langle\ldots\rangle_{\beta}$ denotes
the thermal average with respect to the Galilean measure; i.e., $\left\langle
f[\vec{p}_{\sT}]\right\rangle_{\beta} = \int
d\vec{p}_{\sT}\,f[\vec{p}_{\sT}]\exp{(-\beta
p_{T}^{2}/2 m_k)}$. Actually, we can try to guess what kind of series
we can expect after thermal averaging. It can be easily checked that
for a Galilean particle 
\be\left\langle\lp
\frac{p_{\sT}}{m_k c}\rp^{2n}\right\rangle_{\beta}=
(-1)^{n}\lp\frac{2}{m_{k}c^2}\rp^n
{ d^n\over d\beta^n}\left\langle 1\right\rangle_{\beta}\sim (\beta
m_{k}c^2)^{-n}\left\langle 1 \right\rangle_{\beta}
\ee 
so in the end we will get a series in inverse powers
of $\beta m_{k}c^2$. Notice  that although the coefficients of each
term in the expansion (\ref{t}) are  polynomials in this same
argument, the order of these polynomials is always  bounded by the
corresponding power of $p_{T}/m_kc$ and thus we are only left  with
negative powers of $\beta m_{k}c^2$. The resulting expansion will
then be reliable when $T\ll m_kc^2$ or, in other words, in the limit of
large  masses; this is what one physically should expect, since we
started with  a non-relativistic approximation.

Actually, we shall show that each term in the expansion can be exactly
computed, and after tracing over the longitudinal discrete momentum, it can
be found

\ber
Z_{1}[\beta] & = &
256 \frac{V_9}{\lp 2\pi\rp^9}\, \sum_{k=1}^{+\infty}\,
e^{-\beta m_kc^2}\,\lp\frac{2\pi m_k}{\beta}\rp^{9/2}\times\nonumber\\
& &\times \sum_{n=0}^{+\infty}\frac{\Gamma\lc n+11/2\rc}
{\Gamma\lc n+1\rc \Gamma\lc
-n+11/2\rc}\lp 2\beta m_k c^2\rp^{-n} 
\label{serie} 
\eer

Incidentally, let us mention that it is precisely through this asymptotic
expansion that we see, after identifying $R_{11}$ with
$g_s\sqrt{\alpha'}$, the expected string non-perturbative effects as seen in
a light-cone or infinite momentum frame description, i.e., weighted by the
exponential of minus a constant over a single power of the string coupling.
Looking at the series we see that the
$n=0$ term closely resembles the full low energy DLCQ calculation
(see \cite{ours}).  Here, the term to the 9/2 power comes from the Galilean
kinetic energy of a free particle of mass
$m_k$ in ten dimensions after integrating over the nine-dimensional
(transverse, in eleven dimensions) momentum. The exponential is the
contribution of the rest energy of this particle of mass $m_k$. If one
naively and erroneously identified $R_+$ and $R_{11}$ the difference between
the $n=0$ term in eq.(\ref{serie}) and the whole DLCQ calculation would be
in the counting of the rest energy. It is also important to notice that in
the trace over
$p_{11}$ we have restricted the summation over states with $k\geq 1$ because
we assume that the system has been boosted along the longitudinal direction
to make $p_{11}$ positive. Since in the IMF we have that $p_{11}\gg p_{\sT}$
this means that we will have to take $R_{11}\rightarrow 0$ so the mass of
the Kaluza-Klein states will be very large for a finite $k$. 
This is in agreement with the fact
pointed out above in the sense that (\ref{serie}) is a large mass
$m_{k}c^2\gg T$ (or low transverse velocity $v_{\sT}\ll c$) expansion which
is an asymptotic one over the index $n$ and convergent over $k$ for fixed
$n$. One might expect that taking the
$R_{11}\longrightarrow \infty$ and then summing up on $n$ does not have to
give the same result as summing up the asymptotic expansion and finally
taking the decompactification limit. Furthermore because, as we will see, 
different analytical continuations are involved in each limit.

If we compare with eq.(\ref{Besselg}) and eq.(\ref{asymg}), we notice that
the sum over $n$ reconstructs a Bessel function of $\beta m_{k}c^2$ for a
given $k$. The result is that the single-object partition function can be
written as
\be
Z_{1}[\beta]=2{256\over \beta^{9}}{V_{9}\over (2\pi)^{5}}\sum_{k=1}^{\infty}
(\beta m_{k}c^2)^{5}K_{5}[\beta m_{k}c^2]
\label{series2}
\ee
where $K_{\nu}[z]$ are the modified Bessel functions. In this expression of 
$Z_{1}[\beta]$ we can relax the condition of small $R_{11}$ since the 
series will be convergent for all values of the radius due to the exponential
suppression for large $k$. If we take the limit of large radius in eq. 
(\ref{series2})
we find that the sum can be converted into an integral to give 
(setting again $c=1$)
\be
\lim_{R_{11}\rightarrow \infty} Z_{1}[\beta]= 
2{256\over \beta^{10}}{V_{10}\over (2\pi)^{6}}\int_{0}^{\infty} dx\,x^5\,
K_{5}[x]=256{\Gamma[{11/2}]\over 2\pi^{11\over 2}}V_{10}
\beta^{-10}
\label{half}
\ee 
By simply keeping the zero mode term in the Jacobi $\theta_3$ in
eq.(\ref{proper}) and performing the  integral, it is easy
to check that this result corresponds to one-half the value of the
single-object partition function of a supergraviton in uncompactified
eleven-dimensional space-time. This factor of two difference can be
traced back to the fact that we are just summing over positive momenta
in the $x^{11}$ direction, which corresponds in the computation of the
trace to integrate only over momenta lying on a hemisphere
(i.e. restricting the polar angle on ${\bf S}^{9}$ to the interval
$(0,{\pi\over 2})$). So, we miss half of the degrees of
freedom of eleven dimensional SUGRA. After comparing with
eq.(\ref{properF}) it is also obvious that this factor represents the
anti-D0-brane contribution encoded in the negative KK momentum modes.

From a mathematical point of view and because of its Borel summability,
summing up the asymptotic series would correspond to an analytical
continuation through the Borel transformed series.

The other way around, if we keep $n$ fixed and so take the
decompactification limit then we trail the sum over $k$ by an integral for a
given $n$. More explicitly
\be
\lim_{R_{11}\rightarrow +\infty}\,\sum_{k=1}^{+\infty}\, m_k^{9/2 -
n}\,e^{-\beta m_k} \longrightarrow {{(2\pi
R_{11})}\over{(2\pi)}}\,\int_0^{+\infty}\,dy\,y^{9/2 - n}\,e^{-\beta y}
\ee
to give the following convergent (!) expansion on $n$
\be
Z_1[\beta]= 256{{V_9 (2\pi R_{11})}\over{(2\pi)^{11/2}}}\,
\beta^{-10}\sum_{n=0}^{+\infty}
\frac{\Gamma[n+11/2]}{2^n\,\Gamma[n+1]}
\ee
The convergent sum is easily proven to give
$2^{11/2}\Gamma\left[11/2\right]$. Therefore now we get 
twice the result in eq.(\ref{half}) or, in other words, the complete
supergraviton. Physically this limit corresponds to decompactifying the
Galilean average of each relativistic correction in order to finally sum up
the outputs. Mathematically, the relevant issue is the need for an
analytical continuation for $n \geq 6$ ($n$ is the index that labeled the
asymptotic series and so is zero or a positive integer) of the integral to
which the $k$-sum goes because of the well known restriction upon the
standard integral representation of the gamma function. This is the key
point to understand the different output when computing the
decompactification limit before or after summing up the asymptotic
expansion.

The convergent series we arrive at admits a simple physical
interpretation if we substitute the integral representation for
$\Gamma[n+ 11/2]$ to finally write down a expression in non discrete LCF
variables as

\be
\frac{V_{10}}{\lp 2\pi\rp^{10}}\int_{0}^{+\infty}\,dp^+\,\int\,d\vec{p}_{\sT}
\,\, e^{-\beta p^+}\,
e^{\beta\lp p^+/2 - \vec{p}_{\sT}^{\,2}/2p^+\rp}
\ee
now the transverse momentum is a nine-dimensional vector.

This results from  writing down the energy $p^0$ as $p^+ - p_{L}$ with $p_L
= p^+/2 - p^-/2= p^+/2 - \vec{p}_{\sT}^{\,\,2}/2p^+$. The convergent series
in $n$ comes out as a series for small $p^+$ representing the longitudinal
contribution given by
\be
Z_L[\beta,p^+]= \frac{V_9}{\lp2\pi\rp^9}\,e^{\beta
p^+/2}\,\int\,d\vec{p}_{\sT}\,\,e^{-\beta\vec{p}_{\sT}^{\,\,2}/2p^+}
\ee
Here one must notice that this is the longitudinal contribution in the sense
of the exact light cone frame. 

Through the relationship $s = \beta/p^+$ amongst $p^+$, $\beta$ and the
proper time $s$, this corresponds to an infrared expansion, i.e., one for
large proper times. It is worth remarking that the zero term in this series
coincides with what could be gotten by taking a large longitudinal momentum
and then approaching $p^+$ by $2p_L$. In fact, this convergent series and
the asymptotic one that one gets in eleven dimensions for large longitudinal
momentum (the continuous IMF computation) run very close up to the fifth
term.

Because of its crucial role, it is then  important to show that there exists
a way of getting the asymptotic expansion for the Bessel function as a
result of Watson's Lemma. In the appendix we show how this actually works.

It is through the asymptotic series that one captures all the degrees of
freedom in eleven dimensions. Physically, the starting point is an still
unknown non abelian generalization of the DBI action. One thinks of this
action written as a certain trace over a square root of a non diagonal
matrix. This is a formal writing of the action dictated by a way of making
explicit the ten dimensional relativistic character the action should have
and its scalar character through a certain variation of a trace. In fact,
what we are writing is a Lagrangian with an infinite number of terms are
believed might sum up as a kind of a trace over a square root. The expansion
with an infinite number of terms results as the way of defining the square
root of a non diagonal matrix. Using the corresponding Hamiltonian given by
a series with a finite radius of convergence to compute amplitudes like the
free energy leads to computing momentum integrals which take the momentum to
infinity and then away from the convergence region of the series. These
integrals would finally be (here, we see this in the classical M-theory
limit) Laplace transforms and, as a result of Watson's Lemma, would give an
asymptotic series. This is the way a non abelian DBI action would work as an
effective action and produce asymptotic series and it is in there that the
effective character is encoded. We want to make clear to the reader that we
are not going to propose any modification to any non abelian generalization
of the DBI action.  More than that, we are trying to go further saying that
irrespective of the precise form of such generalization, whenever this is
given by an expansion with a finite radius of convergence, the description
after decompactification one can get for M-theory includes an analytical
continuation performed through the partition function represented as an
asymptotic series.

\section{The effective Lagrangian of the IMF calculation and the 
Dirac-Born-Infeld action}

Recovering the asymptotic expansion of the Bessel function from the IMF
description illustrates more than the way the decompactification limit
should be taken and then eleven dimensional Lorentz covariance would be
achieved. It is clear that the expansion of $p^0$ in relativistic
corrections should be related to an expansion over the Dirac-Born-Infeld
action for D0-branes. In the supersymmetric case, the generalization of this
action to non abelian theories seems to involve the Lagrangian
\cite{9701125}

\be
L= - {c_0\over {g_s}}\, \mbox{Str}\,\sqrt{\det\left(\eta_{\mu\nu} - 
F_{\mu\nu}\right)}
\label{DBIYM}
\ee
together with corrections that cannot be summed up as a symmetric trace (see 
\cite{ProeSevrin}).
Here $\eta_{\mu\nu}=\mbox{diag}\left(1,-1,..,-1\right)$ is the flat Minkowski 
metric in the ten dimensional space where the Dp-brane lives and the 
field strength only depends on the first $p+1$ coordinates. The determinant 
is understood on the matrix character given by the space-time indices
$\mu\nu$. The symmetric trace acts on the $U(N)$ indices and will be understood
by expanding the square root in a multiple power series. Finally $c_0$ is a
normalization constant we take for convenience and $g_s$ is the string
coupling constant.

In the case in which $p=0$ and the group is truncated to $U(1)^N$ 
everything gets easier and one has
\be
L= - {c_0\over {g_s}}\, \mbox{Tr}\sqrt{\left(I-\sum_{i=1}^{9}
\left(\frac{dX_i}{dt}\right)^2\right)}
\ee
with tha gauge election $X_0=A_0=0$, $I$ being the $N\times N$ identity and
$X_i$ is a diagonal $N\times N$ matrix for every $i$ from one to nine. The
standard trace $\mbox{Tr}$ coincides with the symmetric trace when matrices
commute and the corrections to the symmetric trace should also disappear.

Let us now suppose that $k$  elements in the diagonal of $X_i$ coincide for
all $t$. After tracing over, they will sum up to contribute

\be
L_k =  - {c_0\, k\over {g_s}}\,\sqrt{1-\sum_{i=1}^{9}
\left(\frac{dx_i}{dt}\right)^2}
\ee
This is the way we represent the contribution of a bound state of
$k$  D0-branes. We also made the important extra assumption that there is
only one of such states for each $k$ at zero temperature. These are our KK
modes in ten dimensions. To see how to characterize them, let us go to the
Hamiltonian formalism by making a Legendre transform to get

\be
H_k=\sqrt{m_k^2 + p_{\sT}^2},
\ee
where we have already set the constant $c_0$ such that $c_0 k/g_s =
k/R_{11}=m_k$.

The eigenstates of this Hamiltonian will be those of the transverse (in eleven
dimensions) momentum and will also include $k$ as a label of the particle to 
give its mass. It is worth to notice that because of its origin this is a
positive number. 

Computing its single particle partition function $Z_{1,k}$ is a very easy
task. After that it is clear that to get $Z_1(\beta)$ one has to look at $k$
as a quantum number and tracing over it. This clearly enhances the number of
degrees of freedom because it is equivalent to assuming that, at finite
temperature, there is no conservation of the RR charge.
 Another untied question is that of factor 256. In the single particle
calculation in the Hamiltonian formalism it is another quantum number to
trace over to give $2^8$ since for a single particle at finite temperature,
there is no difference between being a fermion or a boson.  It is the spin
quantum number of the quantum mechanics we call Matrix Theory.

With the necessary sum over $k$ from one to infinity, one would get for 
$Z_1(\beta)$ the result in eq.(\ref{series2}) which gives the 'bad' 
decompactification limit. What happens is that we should start from the
theory in interaction (with group $U(k)$) and then, before performing any
kind of trace, one has to use an expansion for the square root what, by the
way, is a tedious task even for the simple case $k=2$. This expansion
includes the rest energy plus the $U(k)$ Super-Yang-Mills contribution to
the Hamiltonian and and infinite number of corrections. The contribution of
this corrections averaged with the Galilean Super-Yang-Mills weight will
finally produce an asymptotic series because of Watson's Lemma. The abelian
limit of this series will be given by eq.(\ref{serie}) through the
asymptotic series of a Bessel function interpreted as an IMF average of the
relativistic corrections.

The key point is then  to explain why, in the case of free D0-branes (the
abelian case), we should prefer the asymptotic series expansion to the
possibility of getting a result without using an expansion for the square
root. The answer arises if one realizes that the abelian $U(1)^k$ case, from
the M-theory point of view, must be understood as a limiting situation on
the non abelian
$U(k)$ picture. So, the way of defining a square root of a non diagonal
matrix is by using an infinite expansion that however has a finite radius of
convergence. Furthermore, trying to write down the adequate generalization
of the DBI Hamiltonian for the non abelian case as some kind of trace of a
square root might simply be a formal task enforced by trying to show the ten
dimensional Lorentz invariance by means of a scalar gotten from a square
root Hamiltonian. The infinite series, when computing amplitudes, will
always give asymptotic expansions by Watson's Lemma. After all, this would
only be an effective action with an infinite number of terms.

We then conjecture that in the $U(N)$ situation one always gets an asymptotic
expansion that is the relevant mean to get the physics, in particular the
decompactification limit will be achieved by decompactifying the
contribution of each term an making an analytic continuation after a
particular order. This order, in the classical limit (free D0-branes in the
Matrix picture), is given by the number of uncompactified dimensions as the
integer part of
$(D+1)/2$ with $D$ the dimension of the open space. It could be that 
D0-brane interactions moved this number down. It is obvious that
compactifying dimensions lowers the order from which the integral
representation of the gamma function fails. To us, the analytical
continuation of the integral that represents the discrete sum for the big
$N$ limit means that the regularization of M-theory Matrix theory
provides includes M-theory effects in the continous limit (the
anti-D0-branes in our computation) only after a kind of extra
regularization over the amplitudes.

\section*{Acknowledgments}
It is a pleasure to thank M. A. V\'azquez-Mozo for very important insights
in an early stage of this work.

\section*{Appendix}
\renewcommand\theequation{A.\arabic{equation}}

That the asymptotic expansion for the  $K_5$ Bessel function can be obtained
by an application of Watson's Lemma seems to be a trivial fact because the
asymptotic series in negative powers of  $\beta m_k c^2$ with coefficient 

\be
c_n =
\frac{\Gamma\lc n+11/2\rc}{2^n\,\Gamma\lc -n+11/2\rc\Gamma\lc n+1\rc}
\ee
is Borel summable and Watson's Lemma is what is behind Borel summability.

However this is not the usual procedure in the mathematical literature to
find the expansion of the modified Bessel functions for large argument. It
seems that this could have been so because things do not appear that trivial
after a first computation since by using the expansion for $p^0$ in powers
of $|p_\sT/m_k c|^2$ (which converges whenever
$|p_\sT/m_kc|< 1$) and averaging with the Galilean Boltzmann factor over
$\vec{p}_{\sT}$ the expansion of $e^{-\beta c p^0}$ in such powers produces for
the single partition function per degree of freedom

\be 
\frac{Z_1[\beta]}{256}=\frac{V_9}{\lp 2\pi\rp^9}\frac{\pi^{9/2}}{\Gamma\lc
9/2\rc}\sum_{k=1}^{+\infty}\,e^{-\beta m_k c^2}
\lp \frac{2 m_k}{\beta}\rp^{9/2} \sum_{n=0}^{+\infty}\,2^n
a_n\lc\beta m_k
c^2\rc\, \Gamma\lc n+9/2\rc \lp \beta m_k c^2\rp^{-n}
\label{watson} 
\ee
where $a_n$ given in eq.(\ref{coefa}) is a polynomial on the variable $\beta
m_k c^2$.
 
In the process of getting eq.(\ref{watson}) one sees that integrating over
$\vec{p}_\sT$ to perform the Galilean average of the relativistic
corrections is finally equivalent to making a Laplace transform. Here we see
the gamma function provided by Watson's Lemma. However we are yet not done
because the coefficients $a_n$ are finite degree polynomials on the variable
$y=\beta m_k c^2$ we must find. If we write them as

\be
a_n\lc y\rc = \sum_{a}b_{n,a}\,y^a
\label{bna}
\ee
one can finally write $Z_1\lc\beta\rc$ per degree of freedom as

\be
\frac{Z_1[\beta]}{256}=\frac{V_9}{\lp 2\pi\rp^9}\sum_{k=1}^{+\infty}
\,e^{-\beta m_k c^2}\lp\frac{2\pi
m_k}{\beta}\rp^{9/2}\sum_{n=0}^{+\infty}c_n \,\,\,(\beta m_k c^2)^{-n}
\ee
with
\be
c_n = \sum_{i=0}^{\infty}\,2^{n+i}\,\frac{\Gamma\lc n+i+9/2\rc}{\Gamma\lc
9/2\rc}b_{n+i,i}
\label{coeficiente}
\ee

To get the coefficients $b_{n+i,i}$ defined in (\ref{bna}) we simply rewrite

\be
e^{-y {\cal C}(x)} = e^y\,e^{yx/2}\,e^{-y\sqrt{1+x}}
\ee
in terms of power series in the dummy variable $x$ ( for $x<1$) and the
variable $y$ to obtain 

\be
b_{n+i,i}= \sum_{z=0}^{i}\sum_{u=0}^{z}\,\frac{(-1)^u}{\lp z-u\rp !\lp
i-z\rp !\,\, u!\,\, 2^{i-z}}\left(
\begin{array}{c}
u / 2 \\
n+z
\end{array}
\rp
\ee

For integer $z,\,i\,$ and $m$ only the odd values of $u$ contribute to the
first sum that can actually be taken up to infinity and summed up. This
gives, after performing the sum over $z$ (generalized hypergeometric
functions help much to perform this task),
\be
b_{n+i,i}=
-\frac{\Gamma[i-n]}{2^{i+2n}\,\Gamma[i]\,\Gamma[1-2n]\,\Gamma[1+i+n]}
\ee

 From here one can easily see that $b_{2n+k,n+k}=0$ with positive integer
$k$ showing that the sum over the index $i$ in eq.(\ref{coeficiente})
actually stops at $n$. There is no problem with setting $i$ to zero, because
$b_{n,0}=0$ for $n\neq 0$ and $b_{0,0}=1$ and the sum over $i$ will start
from $i=1$ for $n > 0$. After substituting this into
eq.(\ref{coeficiente}) one finally gets the expected result for the
coefficient $c_n$.

\end{document}